\documentclass[fleqn,usenatbib]{rasti}
\usepackage{newtxtext,newtxmath}
\usepackage[T1]{fontenc}

\DeclareRobustCommand{\VAN}[3]{#2}
\let\VANthebibliography\thebibliography
\def\thebibliography{\DeclareRobustCommand{\VAN}[3]{##3}\VANthebibliography}

\usepackage{colortbl} 
\usepackage{booktabs} 
\usepackage{graphicx}	
\usepackage{amsmath}	

\title{Mid-IR Water-Vapor Seeing: Implications for ELT/METIS High-Contrast Imaging from VLTI Data}

\author[Raghu, A. S., et al.]{Abhirami S Raghu$^{1}$\thanks{ Corresponding author; abhiramisr24@iitk.ac.in},
Prashant Pathak$^{1}$,
Olivier Absil$^{2}$\thanks{F.R.S.-FNRS Research Director}, 
Benjamin Courtney-Barrer$^{3,8}$, Gilles Orban de Xivry$^{2}$, 
\newauthor Julien Woillez$^{4}$,
Alexis Matter$^{5}$,
Philippe B\'erio$^{5}$,
Roy van Boekel$^{6}$,
Bernhard Brandl$^{7}$
\\
$^{1}$Department of SPASE, Indian Institute of Technology Kanpur, 208016, Uttar Pradesh, India\\
$^{2}$STAR Institute, Universit\'e de Li\`ege, All\'ee du Six Ao\^ut 19c, B-4000 Li\`ege, Belgium\\
$^{3}$Research School of Astronomy and Astrophysics, Australian National University, ACT 2611, Australia\\
$^{4}$European Southern Observatory, Karl-Schwarzschild-Stra\ss e 2, 85748 Garching, Germany\\
$^{5}$Universit\'e C\^ote d’Azur, Observatoire de la C\^ote d’Azur, CNRS, Laboratoire Lagrange, Parc Valrose, B\^at H. Fizeau, 06108 Nice, France\\
$^{6}$Max-Planck-Institut f\"ur Astronomie, K\"onigstuhl 17, Heidelberg 69117, Germany\\
$^{7}$Leiden Observatory, Leiden University, Postbus 9513, 2300 RA Leiden, The Netherlands\\
$^{8}$Australian Astronomical Optics \& School of Mathematical and Physical Sciences, Macquarie University, Sydney 2109, Australia}

\date{Accepted XXX. Received YYY; in original form ZZZ}

\pubyear{\the\year{}}

\begin{document}
\label{firstpage}
\pagerange{\pageref{firstpage}--\pageref{lastpage}}
\maketitle

\begin{abstract}
Direct imaging and characterization of rocky exoplanets remains one of the central challenges in exoplanetary science. The mid-infrared (mid-IR) offers a favorable high-contrast imaging (HCI) regime, as the planet-star contrast is more moderate than at shorter wavelengths. However, mid-IR observations with extremely large ground-based telescopes (ELTs) are strongly affected by spatio-temporal fluctuations in atmospheric water vapor (WV), which introduce chromatic wavefront errors that are largely invisible to adaptive optics systems operating at shorter wavelengths. This poses a significant challenge for mid-IR HCI instruments such as METIS, which aims to detect thermal emission from rocky exoplanets in the $N$-band (8–13 $\mu$m). To quantify WV-induced wavefront errors, we analyze $K$-band fringe-tracker data from the GRAVITY instrument at the VLTI and derive the optical path difference between group-delay and phase-delay measurements. This difference provides a direct measure of the differential WV column density ($\Sigma$) above the telescopes. We examine the dependence of the RMS differential WV column density ($\sigma_{\Sigma}$) on precipitable water vapor (PWV), airmass, seeing, and wind speed across multiple VLTI configurations. Among these parameters, PWV shows the strongest correlation with $\sigma_{\Sigma}$. We further explore the influence of the vertical PWV distribution on this relationship. Next, we use power-spectral-density analysis to quantify how increasing PWV modulates $\Sigma$ variability. The  wavefront error associated with this phenomenon on ELT scale is approximately 15~nm RMS in the $L$-band and 175~nm RMS in the $N$-band, relative to the $K$-band. These results provide a quantitative framework for assessing and mitigating atmospheric WV effects in next-generation mid-IR HCI instruments.

\end{abstract}


\section{Introduction}
High-contrast imaging (HCI) enables the direct detection and characterization of exoplanets and the study of protoplanetary disks \citep{Oppenheimer&Hinkley2009}. These observations are essential for advancing our understanding of planet formation, evolutionary pathways, and the prospects for life beyond Earth. In the mid-infrared (mid-IR), the reduced planet-star contrast significantly enhances the effectiveness of HCI for investigating nearby temperate planetary systems \citep{Pathak2021}. Ground-based mid-IR HCI instruments, such as NEAR \citep[New Earths in the $\mathrm{\alpha}$ Centauri Region,][]{Kasper2019} operating at 10$\ \mathrm{\mu m}$ at the Very Large Telescope (VLT), have demonstrated the ability to detect low-mass companions in nearby stellar systems in the mid-IR. Notably, NEAR has reported evidence for a sub-Saturn temperate planet candidate around $\mathrm{\alpha}$ Centauri \citep{Wagner2021}. Complementary observations with the Mid-Infrared Instrument (MIRI) on the James Webb Space Telescope (JWST) have further underscored the power of mid-IR imaging. The reported detection of a Saturn-mass planet candidate in the habitable zone of $\mathrm{\alpha}$~Centauri~A \citep{Beichman2025}, along with the first direct detection of a temperate super-Jupiter around Eps~Ind~A \citep{Matthews2024}, highlight the emerging capability of mid-IR facilities to probe nearby stellar environments for temperate exoplanets and exozodiacal dust.

\begin{figure}
    \centering
    \includegraphics[width=0.5\textwidth]{{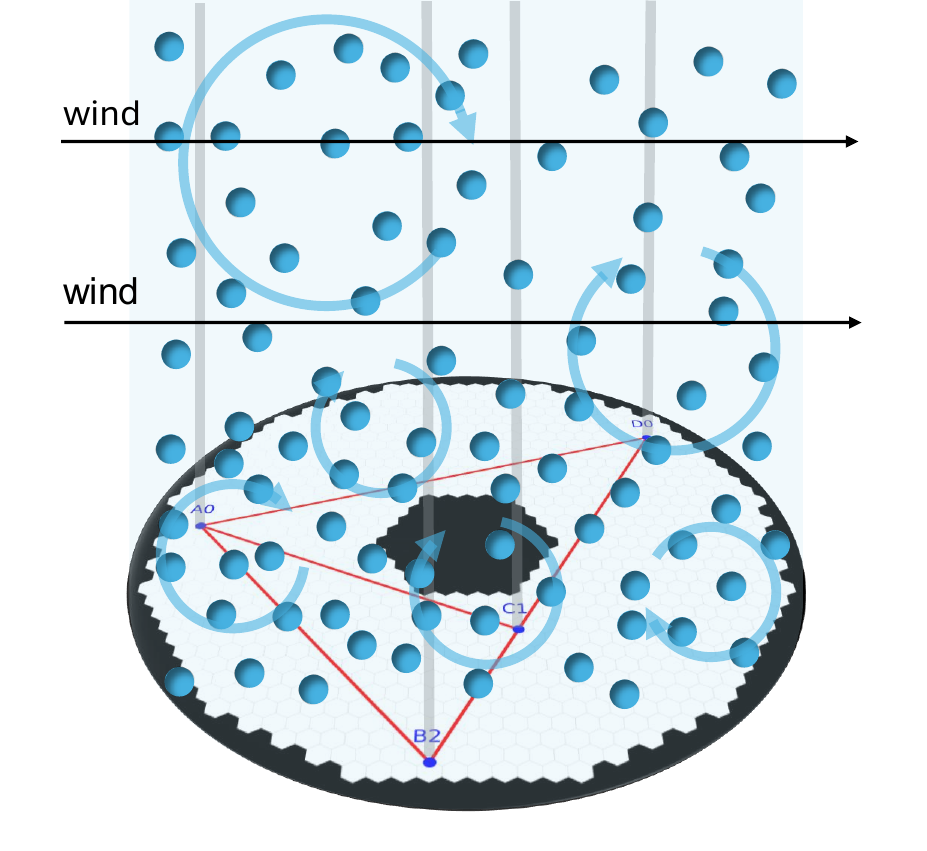}}
    \caption{Conceptual illustration of WV seeing, where blue spheres represent water molecules, and the gray vertical line represent the WV column above the VLTI small AT configuration A0B2D0C1, overlaid on the ELT pupil.}
    \label{fig:pwv}
\end{figure}

In contrast to space-based facilities such as JWST, ground-based telescopes can achieve higher angular resolution, but their performance is fundamentally limited by atmospheric turbulence. In the standard theory of atmospheric turbulence, seeing arises from local refractive-index fluctuations transported by wind and, at optical wavelengths, these fluctuations are predominantly due to dry air. In the mid-IR as well dry-air fluctuations induced-seeing remains the dominant source of wavefront error; however, once this contribution is corrected by an adaptive optics (AO) system, water vapour (WV)-induced turbulence becomes the leading residual term  \citep{Kendrew2008,Absil2022}. WV turbulence, originating from spatial inhomogeneities at various levels in the WV column density are carried across the telescope aperture along the line of sight of the observations by the mean wind flow. Under the frozen-flow (Taylor) hypothesis, the observed temporal evolution of turbulence is driven by advection (wind) rather than by the fluctuations of the inhomogeneous structure. The representative Figure~\ref{fig:pwv} illustrates the distribution of WV molecules in air as blue spheres and the local WV column density as shades of blue across the Extremely Large Telescope (ELT) pupil, overlaid with the small Auxiliary Telescope (AT) configuration of the Very Large Telescope Interferometer (VLTI). It is expected that WV turbulence follows Kolmogorov-von Karman statistics, similar to dry air turbulence, which has been verified experimentally \citep{Lay1997, Masson1994}. 

In the mid-IR, WV affects observations through absorption, increased thermal background, and wavefront distortions. While absorption and background primarily limit sensitivity, this work focuses on wavefront errors, as they degrade image quality and achievable contrast, thereby hindering the detection of faint mid-IR signals such as thermal emission from disks and exoplanets. These WV-induced phase errors arise from the strongly chromatic refractive index of moist air, which exhibits significant wavelength-dependent dispersion across the 8–13~$\mathrm{\mu m}$ atmospheric window \citep{Mathar2004}. As a result, spatial and temporal WV inhomogeneities generate phase fluctuations that differ from those sensed at visible or near infrared (NIR) wavelengths, leaving a chromatic component of turbulence uncorrected by visible/NIR wavefront sensors \citep{Absil2022}. This highlights the importance of understanding WV variability. Unlike homogeneously mixed gases like oxygen, WV exists in all three phases at ambient temperatures and is highly variable both vertically and horizontally. This variability is a function of geographic location, altitude, and time. Precipitable water vapor (PWV) — the equivalent column depth of condensed water (in mm) above the observer — is a common metric utilized to describe atmospheric WV content \citep{Kerber2014}. Therefore, precise site characterization and continuous atmospheric monitoring are vital to subdue the effects of WV on mid-IR observations.

Sites like Cerro Paranal, where the VLT is located, offer excellent conditions for mid-IR observations, with a median PWV of $\sim$2.4~mm  \citep{Querel2014}. The nearby Cerro Armazones site, selected for the ELT, exhibits similar atmospheric characteristics. To support mid-IR observations under such conditions, the European Southern Observatory (ESO) has deployed dedicated atmospheric monitoring instruments. One such instrument is the Low Humidity and Temperature Profiling radiometer \citep[LHATPRO,][]{Querel2014}, which monitors zenith PWV in real time. However, while zenith PWV is routinely monitored and can be specified as an observational constraint, the high spatial and temporal variability of PWV across the sky remains poorly characterized despite its potential to introduce significant errors in HCI observations and long baseline interferometry in the mid-IR. In particular, understanding the effect of PWV on optical path difference (OPD) is pivotal as it is a limiting factor for HCI instruments operating in the mid-IR, such as the Mid-Infrared ELT Imager and Spectrograph \citep[METIS,][]{Brandl2018}, which aims to detect and characterize rocky exoplanets through their thermal signatures in the $N$-band (8–13~$\mathrm{\mu m}$) \citep{Absil2024,Carlomagno2020}. For such instruments, even small variability in wavefront can degrade contrast, directly affecting both the detection of faint planetary signals and the ability to perform high-fidelity spectroscopic characterization of exoplanets \citep{Bowens2021}. Although earlier works \citep{Kerber2014,Querel2014} have conducted a statistical analysis of all-sky PWV homogeneity using the LHATPRO radiometer, it is not sufficient to fully capture the high-frequency temporal fluctuations and small-scale spatial inhomogeneities relevant for HCI. In this study, we aim to fill this gap.

The primary objective of this study is to characterize the impact of spatio-temporal fluctuations in local WV content on wavefront quality for ground-based mid-IR observations. Using archival data from the VLTI, we analyze how fluctuations in WV column density are influenced by atmospheric parameters such as PWV, atmospheric seeing, wind speed, absolute humidity (AH), and relative humidity (RH) across different interferometric configurations, which is pertinent for predicting the performance of future mid-IR instruments such as METIS on the ELT. In Section~\ref{sec:data}, we describe the VLTI/GRAVITY datasets used in this work and the data reduction procedures, together with the methodology for calculating the RMS differential WV column density. We then proceed in Section~\ref{sec:results} with correlating this quantity with atmospheric parameters, analyzing its temporal behavior via power spectral density, and examining vertical humidity and turbulence profiles to assess their impact on the RMS differential WV column density. Our study then derive the impact of WV seeing on mid-IR ELT observations in {Section~\ref{sec:discussion}} , and is finally summarized with concluding remarks in {Section~\ref{sec:conclusion}}.

\section{Observations and data processing}\label{sec:data}
\subsection{Data overview}

\begin{table}
\centering
\begin{tabular}{lcc}
\hline
Configuration & Observations (OBs) & Baselines after filtering  \\
\hline
A0B2D0C1$^{*}$ & 2389 & 1236 \\
A0B2D0I2       &   35 & --   \\
A0B2D0J3       &   38 & --   \\
A0B5D0G1       &   24 & --   \\
A0B5J2J6       &  203 & --   \\
A0G1D0J3       &   12 & --   \\
A0G1J2J3       &  382 & --   \\
A0G1J2K0$^{*}$ & 1421 & 404  \\
A0G2D0J3       &   60 & --   \\
A0G2J2J3       &  217 & --   \\
A0G2J3J2       &   34 & --   \\
K0B2D0J3       &   32 & --   \\
K0G2D0J3$^{*}$ & 2431 & 716  \\
U1U2U3U4       &  496 & --   \\
\hline
\end{tabular}
\caption{Number of OBs (1 OB $\Rightarrow$ 6 baselines) for different VLTI configurations before and after filtering. Configurations used in the final analysis are marked with an asterisk.}
\label{tab:baseline_obs_summary}
\end{table}

\begin{figure}
    \centering
    \includegraphics[width=\linewidth]{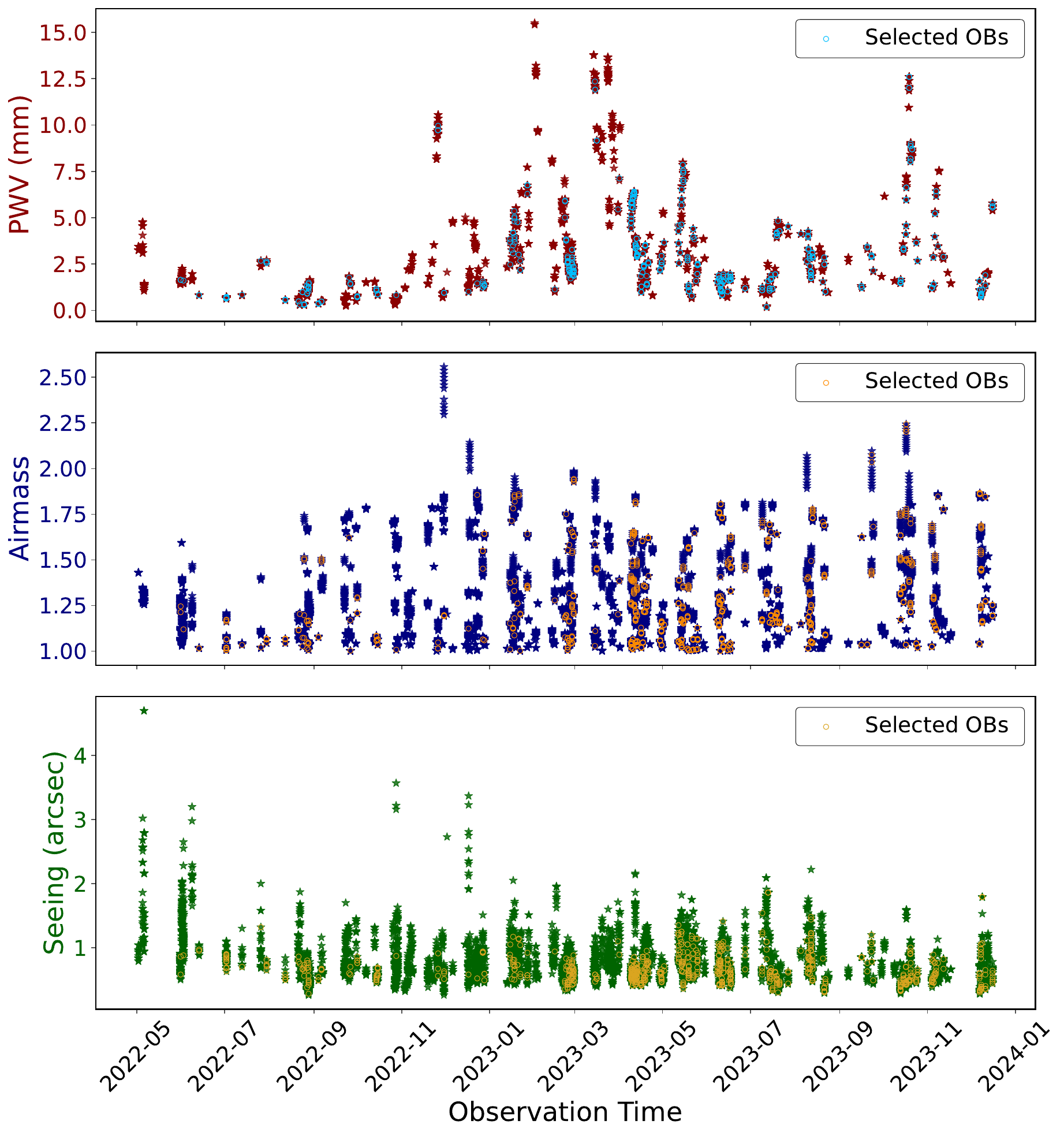}
   \caption{ PWV, airmass, and seeing across observation dates for the small, medium, and large AT configurations are plotted, and the OBs retained after filtering are highlighted with circles.}
    \label{fig:pwv_airmass}
\end{figure}

We analyzed publicly available data from the VLTI/GRAVITY \citep{GRAVITYCollaboration2017} $K$-band fringe tracker recorded during GRA4MAT (GRAVITY {\it for} MATISSE) observations  spanning the period from May 2022 to January 2024. The GRA4MAT mode corresponds to the use of the GRAVITY fringe tracker to stabilize fringes for MATISSE \citep[see][]{Woillez2024}. The dataset covers telescope configurations involving both the 8~m Unit Telescopes (UTs) and the 1.8~m Auxiliary Telescopes (ATs), which offer greater configuration diversity. Table~\ref{tab:baseline_obs_summary} summarizes the configurations along with their corresponding number of observational sequences (OBs). Each OB lasts 60~s and consists of six baselines, recorded with a detector integration time of 1.1~ms. Based on the number of OBs obtained for various configurations, we decided to focus our analysis on the three AT configurations: small (A0B2D0C1), medium (K0G2D0J3), and large (A0G1J2K0)\footnote{\url{https://www.eso.org/sci/facilities/paranal/telescopes/vlti/configuration/P110.html}}. Although UTs also provide a significant number of OBs, we did not include them in our analysis, as they are more prone to vibration-related issues \citep{Courtney-Barrer2022}. Figure~\ref{fig:pwv_airmass} presents the PWV, airmass and seeing conditions for all three selected ATs, demonstrating that the observations provide sufficient coverage across the full parameter range.
The humidity profiles of the Paranal site were obtained from LHATPRO measurements with an integration time of 60~sec, available via the ESO database\footnote{\url{https://archive.eso.org/wdb/wdb/asm/lhatpro_profiles_paranal/form}}. This dataset includes RH and AH measurements at various altitudes, ranging from ground level up to 10,000 meters. Alongside, we retrieved turbulence profiles from the Paranal Multi-Aperture Scintillation Sensor (MASS) and the Differential Image Motion Monitor (DIMM) instruments\footnote{\url{https://archive.eso.org/wdb/wdb/asm/mass_paranal/form}} \citep{Kornilov2007}. DIMM provides integrated seeing measurements at zenith, while MASS resolves the refractive index structure constant $\mathcal{C}_n^2(h)$, representing turbulence strength, in six discrete layers at altitudes of 0.5, 1, 2, 4, 8, and 16~km. Together with the above humidity and turbulence data, wind conditions also contribute to the atmospheric characterization at Paranal. Accordingly, the averaged wind-speed data at 10~m and 30~m altitude were obtained from the Vaisala METEO station at Paranal, accessed via the Paranal Meteorology (METEO) Query Form\footnote{\url{https://archive.eso.org/wdb/wdb/asm/meteo_paranal/form}}.

\subsection{Data processing} \label{sec:processing}
GRA4MAT records the output of the GRAVITY fringe tracker
pixel-to-visibility matrix (P2VM) at each fringe tracker detector integration time in the \texttt{COHERENCE} column of the \texttt{IMAGING\_DATA\_FT} extension \citep{Woillez2024}. This quantity contains the real and imaginary components of the complex coherent flux for the six baselines in four $K$-band spectral channels, together with the corresponding telescope fluxes. For the present analysis, the phase delay (PD) and group delay (GD) were reconstructed offline directly from these complex coherent flux measurements. We emphasize that these quantities are not the temporally filtered signals used by the real-time GRAVITY control loops. In particular, the 150 sample averaging window described by \citet{Woillez2024} applies to the online group delay controller and is not applied in our analysis. Both PD and GD are instead evaluated at every fringe tracker DIT, corresponding nominally to a sampling interval of 1.1~ms, from the same four channel coherent flux measurements.

Let $C_{b,k}(t)$ denote the complex coherent flux on baseline $b$, at time $t$, in spectral channel $k$. The phase delay is estimated from the argument of the coherent sum over the four channels, referenced to its temporal median,
\begin{equation}
    \mathrm{PD}_{b}(t)
    =
    \frac{\overline{\lambda}}{2\pi}
    \arg\left[
        \left(\sum_{k=1}^{4} C_{b,k}(t)\right)
        \widetilde{\left(\sum_{k=1}^{4} C_{b,k}\right)}^{\,*}
    \right],
\end{equation}
where $\overline{\lambda}$ is the mean wavelength, the tilde denotes the temporal median, and the asterisk ($^*$) denotes the complex conjugate. Because the argument of a product with a complex conjugate yields the difference of their phases, this multiplication effectively subtracts the median phase. Adapted from the phase delay formulation in Eq.~11 of \citet{Lacour2019}, this median referencing removes only an arbitrary constant phase origin and does not remove the temporal phase fluctuations of interest.

The group delay is estimated from the phase gradient across the spectral channels using the sum of the adjacent channel cross-spectra,
\begin{equation}
    \mathrm{GD}_{b}(t)
    =
    \frac{1}{2\pi\,\overline{\Delta \sigma}}
    \arg\left[
        \sum_{k=1}^{3}
        C_{b,k}(t) C_{b,k+1}^{*}(t)
    \right],
\end{equation}
where $\sigma=1/\lambda$ is the wavenumber and $\overline{\Delta \sigma}$ is the mean separation between adjacent channels in wavenumber. Both quantities are therefore expressed in units of optical path length.

Following the convention of \citet{Woillez2024}, we define the dispersion metric as
\begin{equation}
    D = \mathrm{PD}-\mathrm{GD}.
    \label{eq:dispersion_metric}
\end{equation}
To illustrate why this quantity remains measurable while the
fringe-tracking loop is closed, the wavelength-independent (i.e., achromatic) optical path may be written as $P_{\rm ach}(t)$, including dry air piston, instrumental piston, and the correction applied by the VLTI delay lines. To first order, the phase and group delays can then be written as
\begin{align}
    \mathrm{PD}
    &=
    P_{\rm ach}
    + N^{w}\Sigma,\\
    \mathrm{GD}
    &=
    P_{\rm ach}
    +
    \left(
        N^{w}
        -
        \lambda\frac{\partial N^{w}}{\partial\lambda}
    \right)\Sigma,
\end{align}
where $\Sigma$ is the differential atmospheric water vapour column density and $N^{w}$ is the specific refractivity of water vapour per molecule. Their difference is therefore
\begin{equation}
    D
    =
    \lambda\frac{\partial N^{w}}{\partial\lambda}\Sigma
    =
    \left(\gamma^{w}\right)^{-1}\Sigma.
    \label{eq:2}
\end{equation}
The common achromatic contribution $P_{\rm ach}$, including the delay-line servo correction, cancels from this expression. Consequently, maintaining the phase delay close to zero does not remove the chromatic water vapour signal from the PD-GD difference.

For the $K$ band, using the water vapour refractivity model of \citet{Colavita2004} adopted by \citet{Woillez2024}, the corresponding group equivalent specific refractivity is
$\left(\gamma^{w}\right)^{-1}\simeq-9.1\times10^{-25}$~cm$^{3}$, or equivalently
$\gamma^{w}\simeq-1.1\times10^{24}$~cm$^{-3}$.
This is the per-molecule form of the $K$-band coefficient tabulated by \citet{Woillez2024}, who give
$\lambda\,\partial N^w/\partial\lambda=-9.13\times10^{-8}$ for their adopted refractivity normalization. The sign records which telescope has the larger water vapour column and has no effect on the RMS or power spectral density quantities used below. The use of the $K$-band PD-GD difference to infer $\Sigma$ was demonstrated on sky by \citet{Woillez2024}, who used it to predict the simultaneously measured MATISSE $N$-band phase.

To exclude intervals in which the GRAVITY fringe tracker did not provide sufficiently stable measurements, we imposed variance thresholds of 1~rad$^{2}$ for GD and 0.1~rad$^{2}$ for PD, corresponding approximately to a $1\sigma$ cutoff. These thresholds are intended to reject intervals affected by poor fringe tracker lock, residual vibrations, or large estimator noise \citep{GRAVITYCollaboration2017}, rather than to select a particular level of atmospheric turbulence. The thresholds were applied uniformly at the level of the complete OB.

\begin{figure}
\centering
\includegraphics[width=0.5\textwidth]{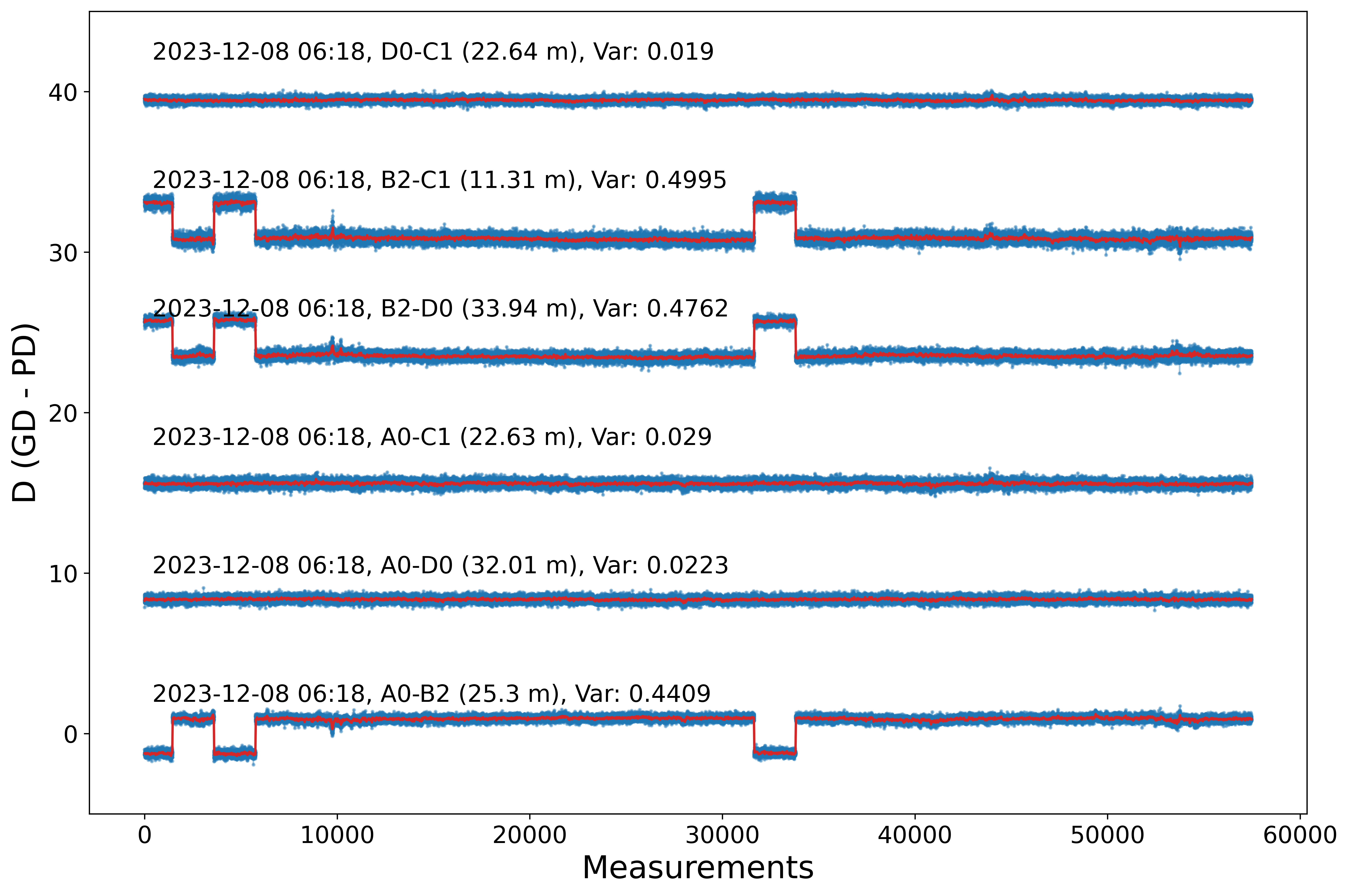}
\caption{Example of raw data of the A0B2D0C1 configuration showing measurements across all baselines for a particular OB showing the dispersion metric $D$. The blue represents the original data, while the red shows the smoothed version.}
\label{fig:before}
\end{figure}

This automated filtering was followed by visual inspection of the dispersion time series to reject OBs containing fringe jumps or short duration discontinuities not identified by the variance threshold. Figure~\ref{fig:before} shows an example containing both valid measurements and intervals affected by fringe jumps and noise. Individual samples with fringe tracker signal-to-noise ratio (SNR) below 10 were also excluded. The smoothing shown in Fig.~\ref{fig:before} is used only as a visual guide; all RMS and power-spectral-density calculations use the unsmoothed, frame level measurements.

A summary of the baselines retained after filtering is provided in Table~\ref{tab:baseline_obs_summary}. After filtering, 1236 OB sequences remained for A0B2D0C1, corresponding to 8.62\% of the 2389 OBs multiplied by six baselines. The equivalent numbers were 404 sequences for A0G1J2K0 (4.74\%) and 716 for K0G2D0J3 (4.91\%). The retained observations continue to span the
full range of atmospheric conditions, as shown in Fig.~\ref{fig:pwv_airmass}. The retained dispersion sequences were converted to differential water vapour column density using Eq.~\ref{eq:2}. For each OB sequence, we computed the standard deviation of the unsmoothed $\Sigma(t)$ measurements to obtain the RMS differential column density $\sigma_{\Sigma}$. The arbitrary constant introduced by the temporal phase reference does not affect this standard deviation. It is customary to express $\sigma_{\Sigma}$ as an equivalent thickness of precipitable water. A column-density RMS of $10^{19}$~cm$^{-2}$ corresponds approximately to a water thickness of 3~$\mathrm{\mu m}$ \citep{Absil2022}. We denote this equivalent RMS thickness by $\sigma_{\rm wv}$. It provides the principal empirical measure of water vapour seeing used in this study and can be converted into the corresponding chromatic wavefront error as described in Sect.~\ref{sec:statistics}. No explicit subtraction of the PD/GD estimator-noise variance is performed. The measured $\sigma_{\rm wv}$ should therefore be interpreted as the RMS of the retained dispersion signal, including any residual measurement-noise contribution. We assess the importance of this contribution through the SNR selection and the subsequent analysis of the dependence of $\sigma_{\rm wv}$ on fringe tracker SNR and target flux.
\section{Results}
\label{sec:results}

\begin{figure}
    \centering
    
     \includegraphics[width=\linewidth]{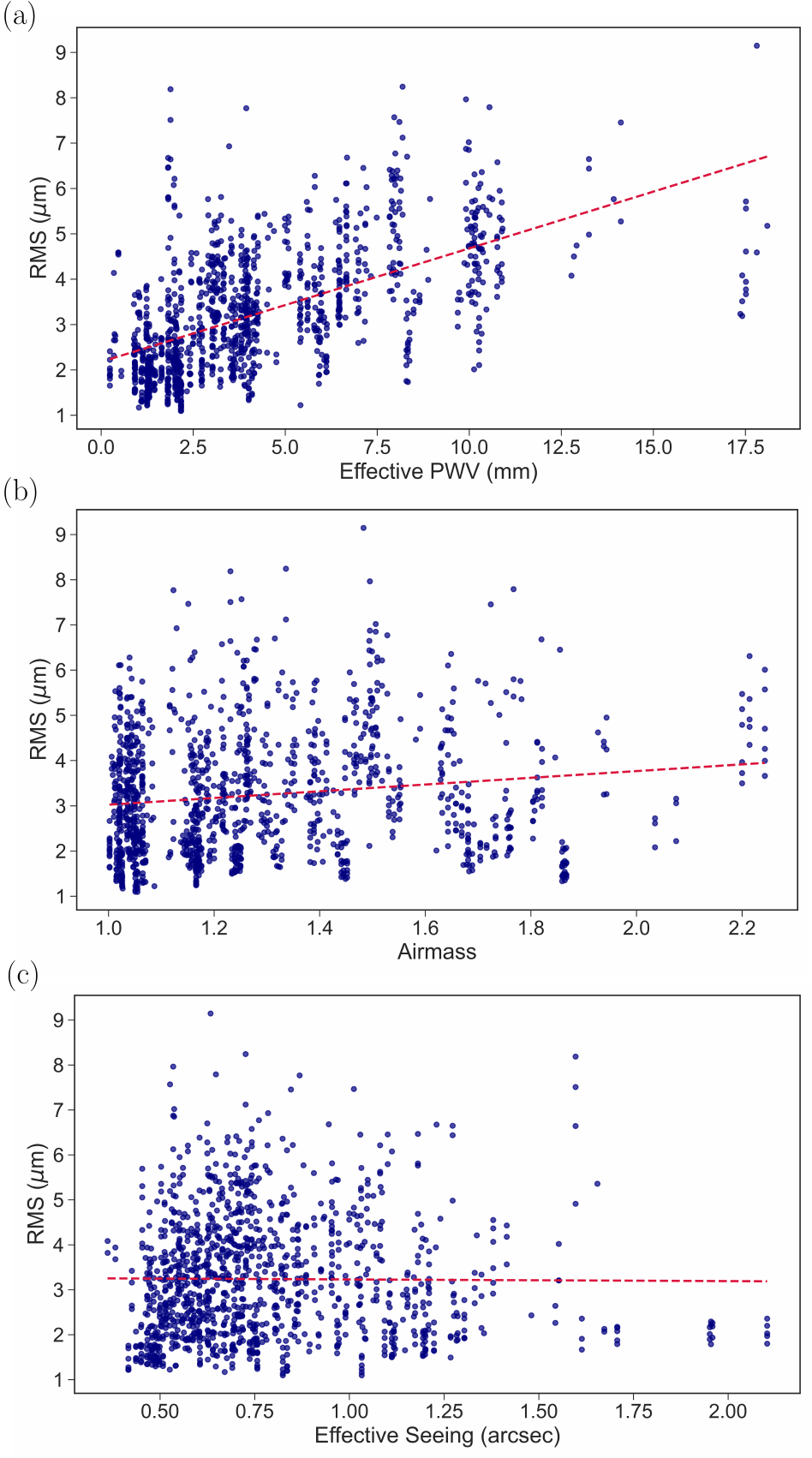}
   
    \caption{Correlation of RMS WV seeing $\sigma_{\rm {wv}}$  with (panel (a)) effective PWV, (panel (b)) airmass, and (panel (c)) effective seeing with linear fit for configuration A0B2D0C1.}
    \label{fig:rms_three_plots}
\end{figure}

\subsection{Influence of environmental parameters on WV seeing}

Before considering the effect of instrumental parameters, we start by exploring the correlation of $\sigma_{\rm {wv}}$ fluctuations with atmospheric parameters such as PWV (effective), seeing (effective) and airmass. Figure~\ref{fig:rms_three_plots} shows the correlation of $\sigma_{\rm {wv}}$ with effective PWV, airmass, and effective seeing for the small configuration A0B2D0C1. Panel (a) shows a strong correlation between $\sigma_{\rm {wv}}$ and effective PWV, indicating that higher mean PWV corresponding to an increased level of WV, provides room for larger $\sigma_{\rm {wv}}$ fluctuations, whereas no significant correlation is observed with either airmass (panel (b)) or effective seeing (panel (c)).

\begin{figure}
    \centering
        \includegraphics[width=\linewidth]{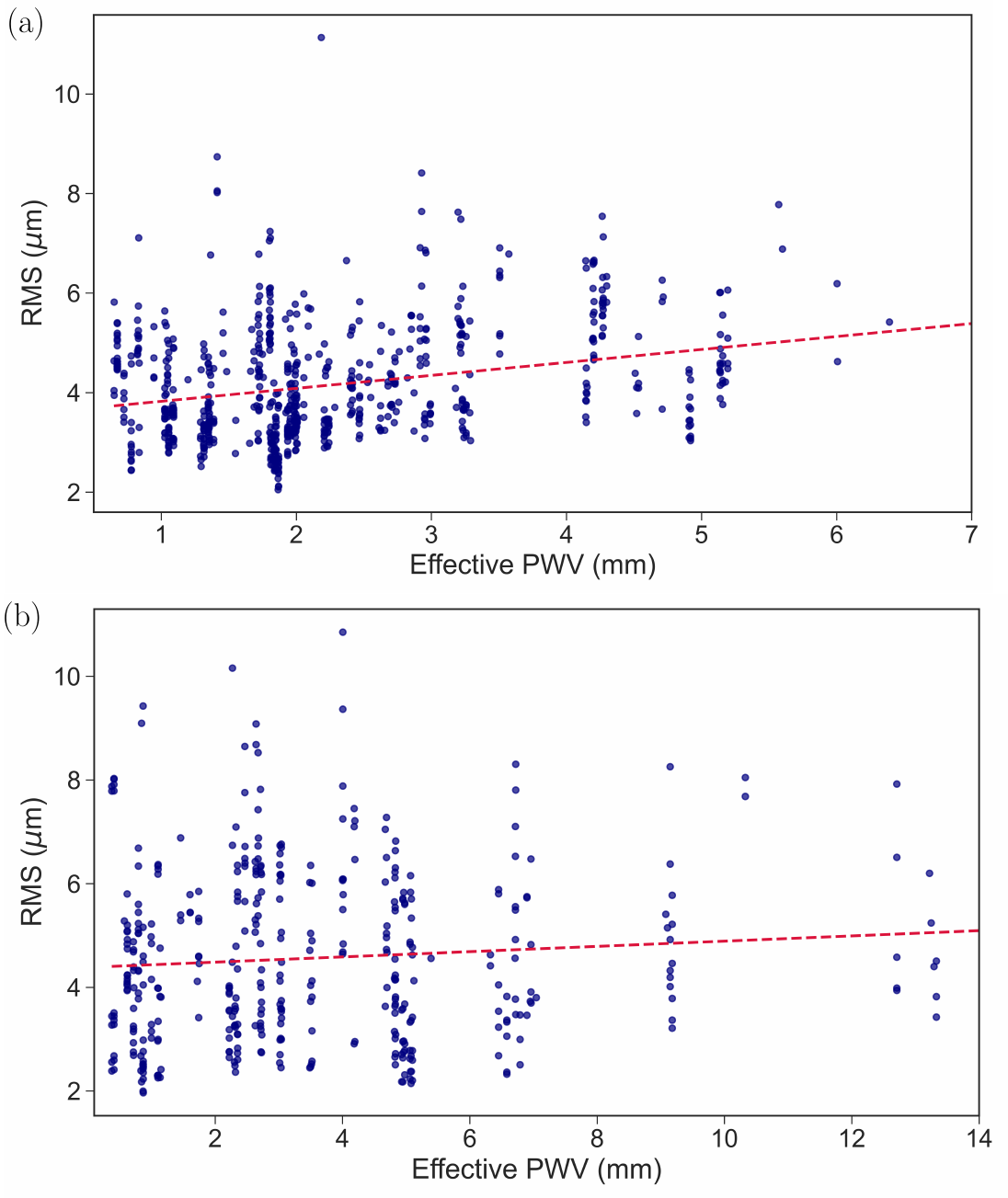}
    \caption{$\sigma_{\rm {wv}}$ variation with effective PWV for medium (K0G2D0J3, panel (a)) and large (A0G1J2K0, panel (b)) configurations.}
    \label{fig:RMSvariation}
\end{figure}

Given the strong correlation of WV seeing with effective PWV, we continue our analysis by exploring the correlation between $\sigma_{\rm {wv}}$ and effective PWV across medium and large configurations. The medium configuration exhibits a weaker correlation, and the large configuration displays no significant correlation, as illustrated in Fig.~\ref{fig:RMSvariation}. We believe that this trend arises because, at longer baselines, we observed lower fringe tracker SNR, leading to higher noise levels and fewer statistically usable data points, which tends to conceal a possible underlying correlation between $\sigma_{\rm {wv}}$ and effective PWV in the medium and large configurations. Given that A0B2D0C1 not only shows the strongest correlation but also because its baseline lengths are more relevant to the spatial scales probed by the ELT pupil, we focus on this configuration for further analysis. We also examined the $\sigma_{\rm {wv}}$–effective PWV correlation for each individual baseline within this AT configuration to determine whether baseline geometry affects the observed behaviour, and found no significant dependence on baseline length nor orientation within our statistical uncertainties. Therefore, all baselines from A0B2D0C1 are considered jointly in the rest of the analysis.

Despite the strong correlation between $\sigma_{\rm wv}$ and PWV, the relationship exhibits substantial scatter about the linear trend. To investigate whether atmospheric turbulence quantified by the dry-air seeing contributes to this dispersion, we examined the dependence of $\sigma_{\rm wv}$ on seeing by binning the data according to standard ESO seeing categories. Specifically, we adopted the commonly used thresholds of $<0\farcs6$, $<0\farcs8$, $<1\farcs0$, and $>1\farcs0$, corresponding to approximately $20\%$, $50\%$, $70\%$, and $100\%$ of seeing occurrences, respectively. The dataset was therefore divided into four bins as defined above. We find no significant dependence of $\sigma_{\rm wv}$ on seeing: a comparably strong correlation between $\sigma_{\rm wv}$ and effective PWV is present across both good and moderate seeing conditions, without any reduction of the scatter in the data points around the linear trend. We further analyzed whether the locally averaged wind speed could account for the observed dispersion by examining the correlation of wind speed at 10~m and 30~m with $\sigma_{\rm {wv}}$, as well as with PWV measurements. However, no significant correlation was observed.

Next, we analyzed the correlation between the spatial structure of atmospheric WV and $\sigma_{\rm wv}$. We estimated the spatial de-correlation scale of WV ($L_{\rm pwv}$) using PWV measurements from LHATPRO with an integration time of $5$~s. The temporal de-correlation of PWV fluctuations was estimated using the autocorrelation function (ACF). We define the temporal de-correlation time ($\tau_{1/e}$) as the time lag at which the normalized ACF decreases to $1/e$ of its peak value. The $\tau_{1/e}$ was then converted to $L_{\rm pwv}$ using the Taylor frozen-flow approximation, $L_{\rm pwv}=v\times\tau_{1/e}$, where $v$ is the local wind speed. The resulting PWV spatial de-correlation scales are of the order of a few hundred meters to a few km, much larger than for dry air seeing as already reported in the literature \citep{Lay1997}. 
We used a time window of 10-minute and 1-hour around each $\sigma_{\rm wv}$ measurement, based on the available corresponding PWV timeseries, we find no correlation between $L_{\rm pwv}$ and the measured $\sigma_{\rm wv}$.

\begin{figure}
  \centering
  \includegraphics[width=\linewidth]{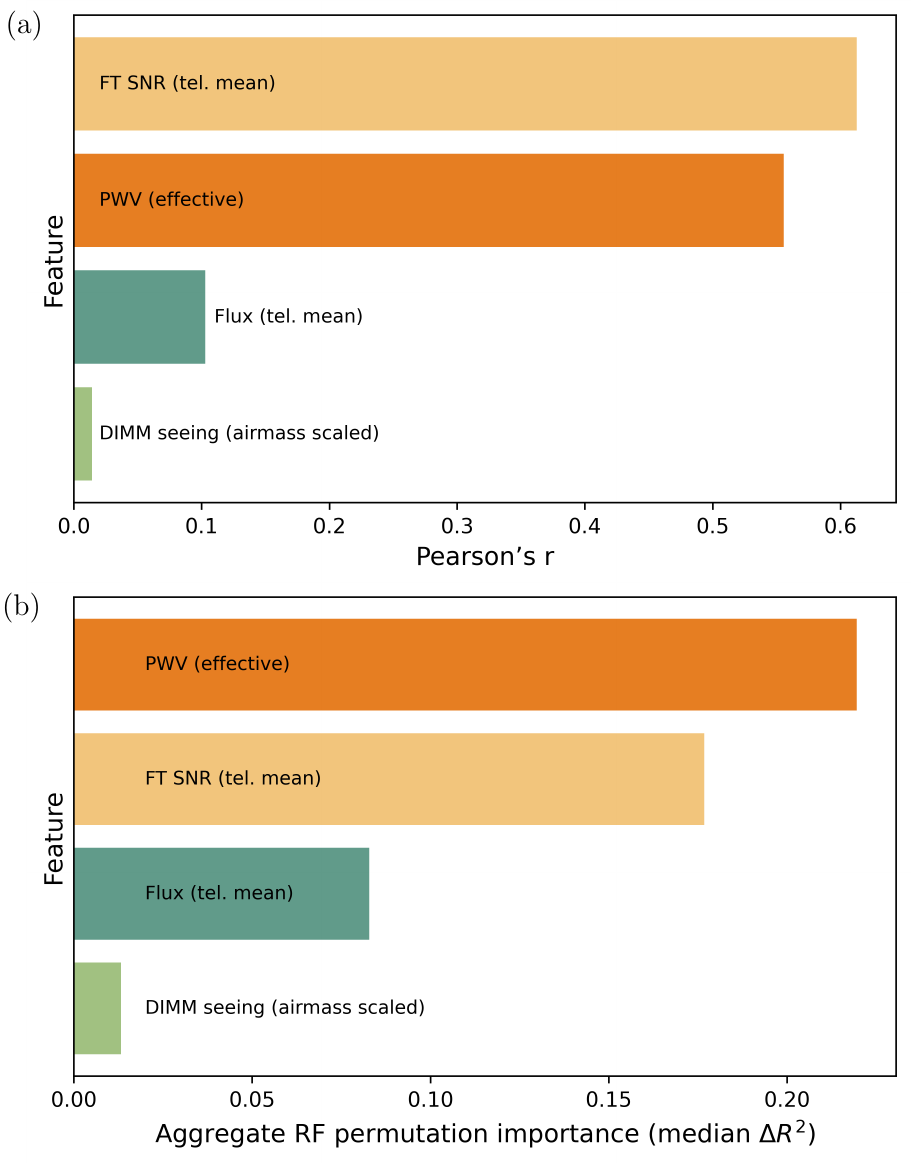}
  \caption{Feature importance analysis, showing the marginal Pearson’s $r$ versus conditional signal (panel (a)), and RF permutation importance (panel (b)) for the selected predictors. Predictor labels that include ``(tel.\ mean)'' indicate that the underlying per-telescope channels were first averaged across all contributing telescopes for each exposure, so the plotted importance reflects the influence of these telescope-averaged predictors.}
  \label{fig:pwv_humidity_ML}
\end{figure}
To further explore the impact of atmospheric parameters on $\sigma_{\rm {wv}}$, while also considering fringe tracker SNR and star brightness as potential contributors to WV seeing, we use a feature importance analysis, where we assess the contribution of various predictors $X_j$ (e.g. seeing, PWV, etc.) to the target $Y=\sigma_{\rm {wv}}$ using two complementary methods. The top panel of Fig.~\ref{fig:pwv_humidity_ML} shows the marginal linear signal, quantified by the absolute Pearson correlation 
$\ |r_j| = |\mathrm{cov}(X_j,Y)/(\sigma_{X_j}\sigma_Y)|$, which measures the direct linear relationship between each predictor and  $\sigma_{\rm {wv}}$. The bottom panel show the conditional contribution of each predictor $X_j$ to $\sigma_{\rm {wv}}$. We compute permutation importance from a random-forest (RF) regressor trained separately for each baseline $b$. For a given baseline model, let $R_b^2$ denote its (coefficient of determination) performance when evaluated on the unperturbed data. To assess the influence of a specific predictor $X_j$, we randomly permute the values of $X_j$ within that baseline—thus destroying any relationship between $X_j$ and $\sigma_{\rm {wv}}$—and recompute the model performance. We denote the resulting score by $R_{b,j}^{2\,(\pi)}$. The 
permutation importance for feature $X_j$ on baseline $b$ is then
\begin{equation}
\Delta R^{2}_{b,j}
  = R_b^2
  - \mathbb{E}_{\pi}\!\left[\,R_{b,j}^{2\,(\pi)}\,\right],
\end{equation}
where $\mathbb{E}_{\pi}[\cdot]$ denotes the average over repeated random permutations $\pi$. A larger $\Delta R^{2}_{b,j}$ indicates that shuffling $X_j$ leads to a substantial drop in predictive performance, and hence that $X_j$ is informative for that baseline. Because each baseline is analysed independently, we obtain one importance value \(\Delta R^{2}_{b,j}\) per baseline \(b\). To provide a single, robust instrument-wide ranking of predictor relevance, we aggregate the baseline-wise importances by taking the median across all baselines. Not surprisingly, the fringe tracker SNR turns out to have a very strong contribution, of the same order as PWV, which may (at least partly) explain the large scatter in Figs.~\ref{fig:rms_three_plots} and \ref{fig:RMSvariation}, especially at the longest baselines where the fringe tracker SNR is lower. Star brightness also has a non-negligible influence and may also add to the scatter in said figures.

\subsection{Influence of vertical humidity and turbulence profiles}
The strength of atmospheric turbulence, captured by the structure constant $\mathcal{C}_n^2$, is known to vary strongly with altitude. Since WV seeing is expected to be caused by the same wind-driven processes as dry air seeing, it is interesting to explore how the combination of the $\mathcal{C}_n^2$ profile and the variation in WV content across the atmosphere may affect WV seeing. Here, we start by examining the correlation of PWV with AH and RH at different atmospheric heights, as PWV serves as a proxy for $\sigma_{\rm {wv}}$. PWV correlates positively with AH in all layers. RH, which depends on temperature, does not show a significant correlation with PWV at any altitude.
\begin{figure}
    \centering
    \includegraphics[width=\linewidth]{ 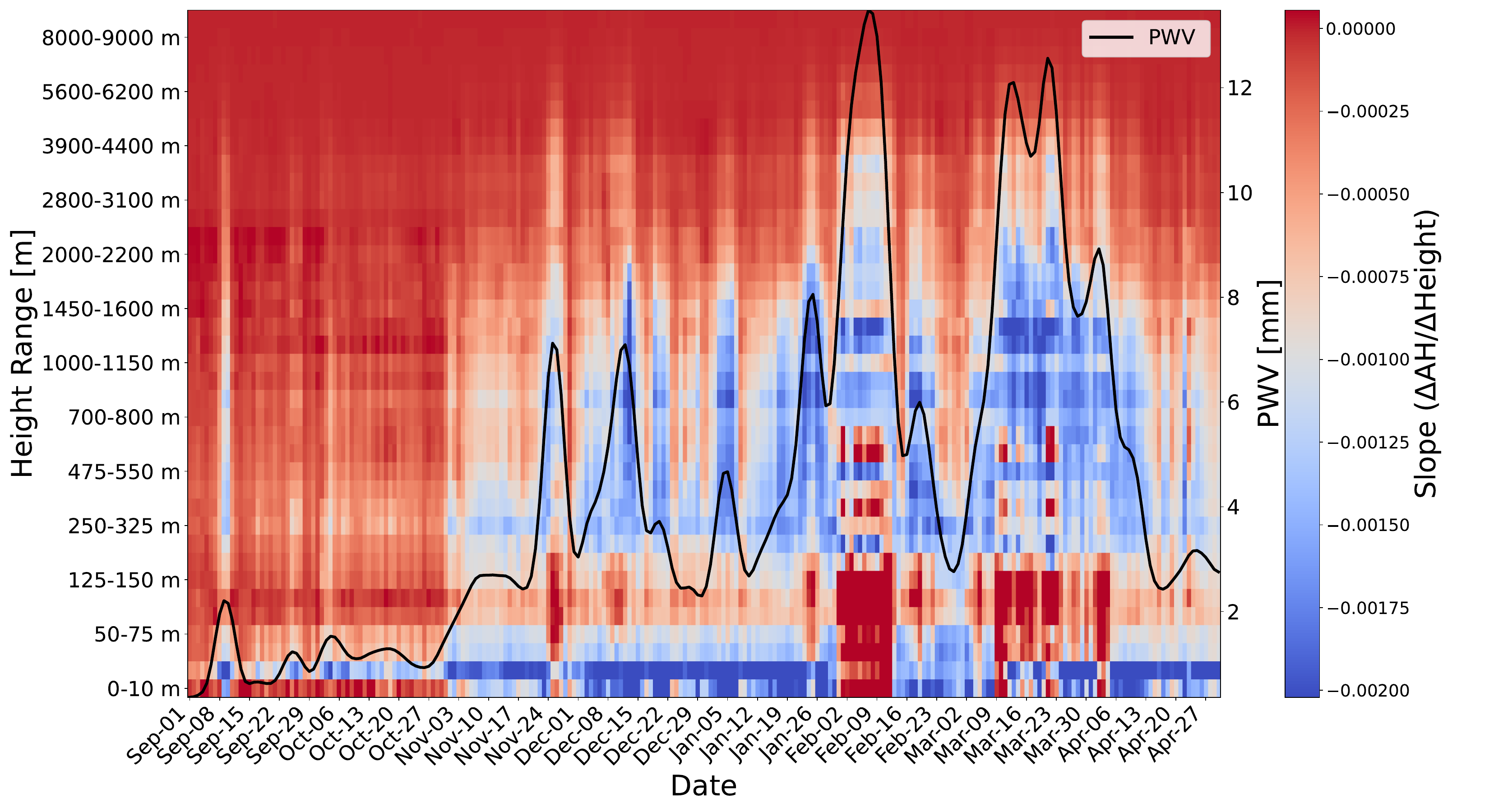}
    \caption{Time evolution of the vertical gradient of AH across different height layers is shown. The black line overlaid indicates the corresponding PWV in mm.}
    \label{fig:ah_gradient_pwv}
\end{figure}
To further investigate the role of the AH distribution across the atmosphere in the PWV variability, we analyzed the temporal evolution over a year of the local vertical gradient of AH (i.e., $\Delta$AH/$\Delta$Height) across atmospheric layers from the surface up to 10~km. Figure~\ref{fig:ah_gradient_pwv} shows the results, where the color scale represents the AH gradient as a function of time and height, and the overplotted black line indicates the PWV measurements (data was taken from the LHATPRO). This highlights that PWV variability is strongly modulated by vertical AH gradients in the lower (<2~km) and mid-troposphere (2–8~km), which in turns suggests that PWV variability is mostly driven by low- to mid-altitude atmospheric layers.

To further understand how atmospheric turbulence affects the $\sigma_{\rm {wv}}$ variations, we analyze turbulence profiles obtained from MASS and DIMM observations. MASS data provide vertical $\mathcal{C}_n^2(h)$ turbulence profiles across six layers, while DIMM gives integrated seeing measurements. We also have a Layer-0 turbulence estimate that represents ground-layer contributions, obtained by subtracting MASS from DIMM data. 
\begin{figure}
   \centering
  \includegraphics[width=\linewidth]{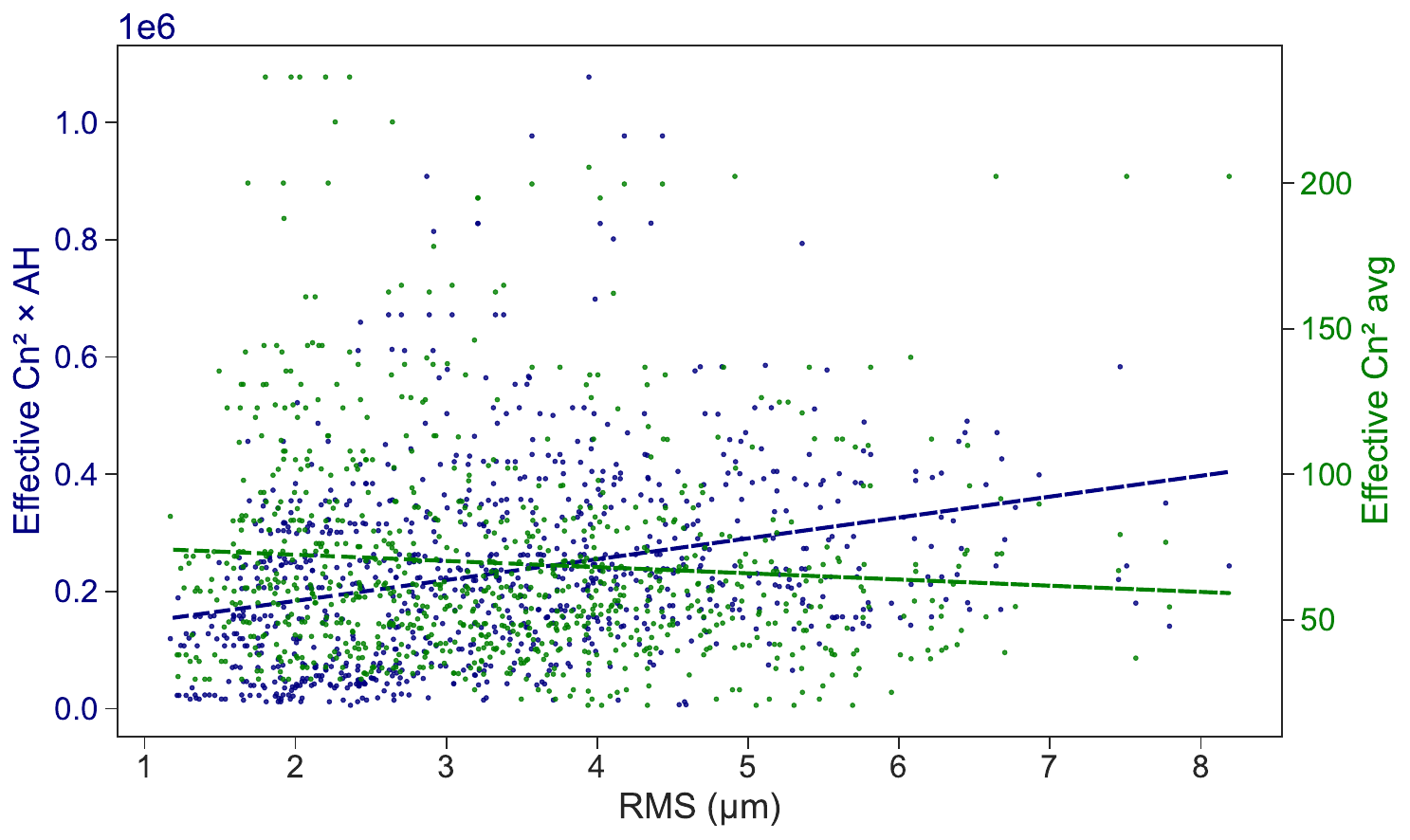}
  \caption{ $\sigma_{\rm {wv}}$ against AH-weighted Effective~$\mathcal{C}_n^2 $(blue) and averaged effective $\mathcal{C}_n^2$ (green).}
\label{fig:cn2_rms}
\end{figure}
As shown in Fig.~\ref{fig:cn2_rms}, $\sigma_{\rm {wv}}$ variations exhibit a more significant linear trend with the AH-weighted effective $\mathcal{C}_n^2$ compared to the effective turbulence strength (effective $\mathcal{C}_n^2$) alone. These AH and turbulence profiles suggest that the primary source of $\sigma_{\rm {wv}}$ variations is the lower atmosphere, where both turbulence and humidity are concentrated.

\subsection{Power spectral density analysis}

To explore the frequency-dependent behaviour of WV fluctuations, we computed the PSD of $\Sigma$ for different PWV conditions.
A fixed-frequency grid, ranging from 0.017~Hz to 454~Hz, was used to accommodate OBs with observation time close to $60$~sec. For this, we employ the Lomb-Scargle periodogram, widely applied in astrophysical time series analysis and particularly effective in addressing any possible irregularities in temporal sampling. For comparison, the PSDs of $\Sigma$ were grouped into five 3~mm-wide PWV bins spanning 0–15~mm, and the PSDs within each bin were averaged to determine the mean PWV and PSD for that bin.
\begin{figure}
	\centering
	\includegraphics[width=\linewidth]{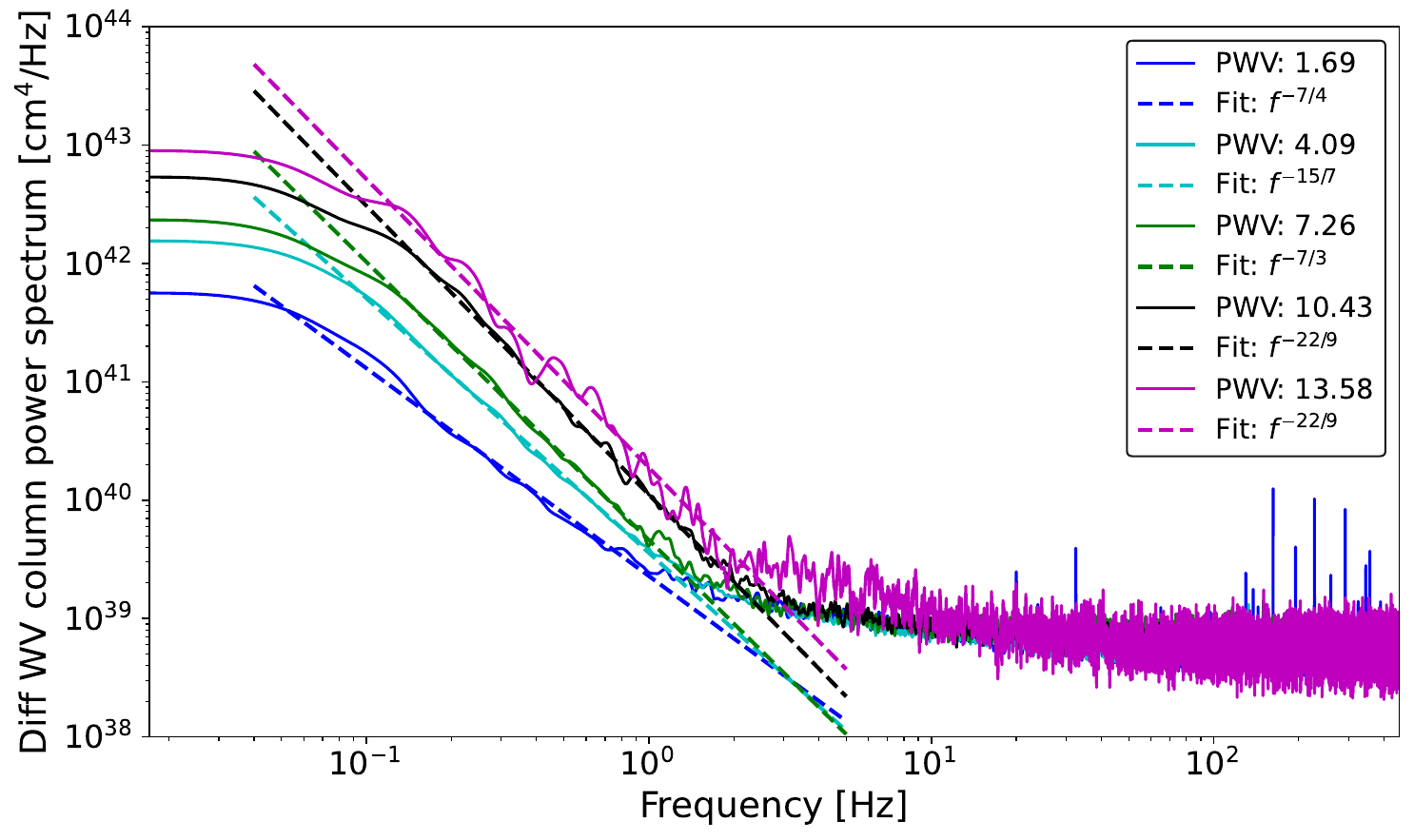}
	\caption{PSD for $\Sigma$ fluctuations in different PWV conditions across the baseline A0-B2 (25.30~m) for the A0B2D0C1 configuration.}
	\label{fig:psd_pwv_A0B2D0C1_detailed}
\end{figure}
The PSDs for baseline A0–B2 (25.3~m) are shown in Fig.~\ref{fig:psd_pwv_A0B2D0C1_detailed}, where each curve corresponds to a different PWV bin, and the dashed lines represent a power-law fit to the PSDs in log–log space, performed over the 0.1–1~Hz temporal-frequency range, which provides an estimate of the slope of the spectral power distribution ranging from $-1.8$ at low PWV to $-2.4$ at high PWV.

The analysis reveals that low-frequency wavefront fluctuations are enhanced under high-PWV conditions, which could affect long-exposure mid-IR interferometric imaging \citep{Colavita2004}, such as in the MATISSE \citep{Woillez2024} and ASGARD/NOTT \citep{Laugier2024} instruments of the VLTI.
The expected slope of the PSD for $\Sigma$ is directly linked to that of the differential piston, which is theoretically predicted to be $-8/3$ under Kolmogorov turbulence theory \citep{Conan1995}. Our analysis yields PSD slopes that are consistent with theoretical value at high PWV, while at low PWV a shallower slope is observed. This behavior is consistent with \citet{Colavita2010}, where the median WV power spectra follow the Kolmogorov $-8/3$ slope, while the ensemble-average spectra exhibit slightly shallower 
slope. These results implies that the turbulence-driven behavior becomes more evident as the atmospheric WV content increases, and a similar trend is also observed for longer baselines.

\subsection{WV seeing statistics} \label{sec:statistics}

\begin{figure}
	\centering
	\includegraphics[width=\linewidth]{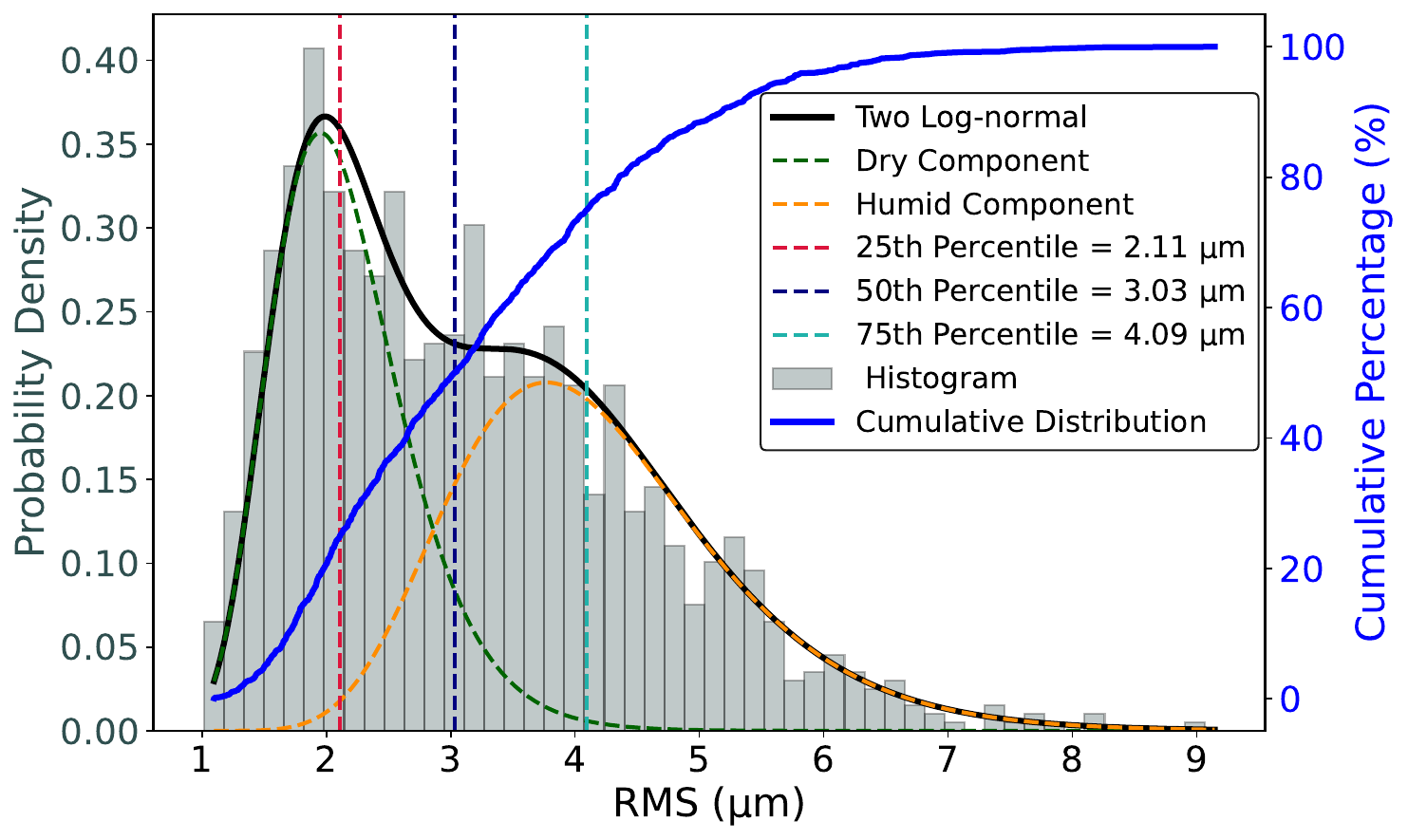}
	\caption{Histogram and cumulative distribution of $\sigma_{\rm {wv}}$ values. The grey bars represent the frequency distribution of $\sigma_{\rm {wv}}$, while the blue curve shows the cumulative percentage. The black curve denotes the two-log-normal fit, with the green and orange dashed curves representing the dry and humid components, respectively. The dashed vertical lines indicate the 25th(red), 50th (median) (navy blue), and 75th (cyan) percentiles, which are $2.11~\mathrm{\mu m}$, $3.03~\mathrm{\mu m}$, and $4.09~\mathrm{\mu m}$, respectively.}
	\label{fig:cdf_rms_A0B2D0C1}
\end{figure}

Having identified the most suitable interferometric configuration for deriving WV seeing and examined its behaviour under a range of conditions, we now turn to the arguably most significant outcome of our analysis: the statistical occurrence of different WV-seeing regimes. Quantifying these occurrence rates is crucial to understand the typical wavefront perturbations that will be encountered by future mid-IR instruments like METIS. Here, we analyze the statistical distribution of $\sigma_{\rm {wv}}$ using histograms and cumulative distribution functions (CDFs). These distributions capture the typical scale and spread of WV seeing, as shown in Fig.~\ref{fig:cdf_rms_A0B2D0C1}. The histogram (grey bars) shows the frequency of $\sigma_{\rm {wv}}$ values, while the overlaid CDF (blue curve) represents the cumulative percentage of occurrences. Vertical dashed lines indicate the 25th, 50th (median), and 75th percentiles, highlighting the central tendency and variability of $\sigma_{\rm {wv}}$ in the small A0B2D0C1 configuration. From the figure, it can be statistically inferred that most values of $\sigma_{\rm {wv}}$ lie between approximately 2 and  $4~\mathrm{\mu m}$, with a median around  $3~\mathrm{\mu m}$, and remain below  $6~\mathrm{\mu m}$ \citep[which was the value originally considered in the study of][]{Absil2022} about 95\% of the time. In addition, it seems that the histogram may be described by the sum of two log-normal distributions, which might respectively correspond to relatively dry conditions associated with lower $\sigma_{\rm {wv}}$ values and and more humid conditions associated larger WV induced fluctuations.
\section{Impact on mid-IR ELT observations} \label{sec:discussion}
In order to understand the effect of the median level of WV seeing ($\sigma_{\rm {wv}}=3~\mathrm{\mu m}$) on mid-infrared adaptive optics observations, we derive the wavefront error from $\sigma_{\Sigma}$, which directly relates to $\sigma_{\rm {wv}}$ (see Section~\ref{sec:processing}). For that, we first need to convert this quantity into an RMS optical path difference $\sigma_{\rm {OPD}}$ at the wavelength of interest, relative to the wavelength at which the wavefront is sensed and controlled by the adaptive optics system. This conversion is based on the concept of water displacing air \citep[WDA,][]{Meisner&Poole2003}, where fluctuations in the WV column density, under constant pressure and temperature, lead to opposite variations in the dry air column density. The specific refractivity of WDA (in cm$^3$) is then given by:
\begin{equation}
   \ N_{\rm WDA} = N_{\rm w} - N_{\rm air} \, .
\end{equation}
The chromatic path difference created by a $\sigma_{\Sigma}$ of PWV is then simply given by $N_\mathrm{WDA}\times \sigma_{\Sigma}$, in cm. 

Our median level of WV seeing of about $3~\mathrm{\mu m}$ RMS corresponds to an equivalent $\sigma_{\Sigma}$ of $10^{19}~\mathrm{cm}^{-2}$. We can then convert this RMS column density into an equivalent differential OPD between standard infrared atmospheric windows. More precisely, we use the model of \cite{Mathar2004} to compute the refractivity of WV in the middle of the $K$, $L$, and $N$ atmospheric windows \citep[see][for more details]{Absil2022}, and get the following values for the differential specific refractivity of WDA: $N_\mathrm{WDA}(L-K)=2.5\times 10^{-25} \, \mathrm{cm}^3$ (per molecule), and $N_\mathrm{WDA}(N-K)=3\times 10^{-24} \, \mathrm{cm}^3$ (per molecule). This results in RMS optical path differences ($\sigma_{\rm OPD}$) of 25~nm (between $K$ and $L$ bands) and 300~nm (between $K$ and $N$ bands). 

Following \cite{Absil2022}, we can further convert this (temporal) $\sigma_{\rm {OPD}}$ into an equivalent (spatial) RMS wavefront error at the ELT scale, recognizing that the VLTI compact configuration (A0B2D0C1) has the exact same scale as the ELT pupil, and assuming a Kolmogorov-von Karman spatial power spectrum for WV turbulence. This results in (spatial) RMS wavefront errors of about 15~nm at $L$-band relative to $K$-band, and of about 175~nm at $N$-band relative to $K$-band. Behind adaptive optics correction at $K$-band, which provides an expected residual wavefront error of typically 128~nm RMS in an instrument like METIS \citep{Feldt2024}, WV seeing is therefore bound to become the dominant source of wavefront error at $N$-band, especially for low spatial frequencies where most of the Kolmogorov-von Karman power resides (while adaptive optics residuals have a low power content at low spatial frequencies).

Consistent with this estimate, METIS performance assessments indicate that WV turbulence will be a major limitation for mid-IR HCI. End-to-end High-contrast End-to-End Performance Simulator (HEEPS) studies \citep{Delacroix2022} show that realistic WV phase screens substantially degrade coronagraphic performance across the $L$–$M$–$N$ bands, with the $N$ band exhibiting the strongest contrast degradation. An updated and more detailed analysis of the impact of WV seeing on HCI performance is presented in our recent work~\citep{Absil2026}. System-level analyses of the METIS single-conjugate adaptive optics (SCAO) architecture \citep{Feldt2024} similarly identify WV-induced chromatic phase errors as a dominant residual term that remains uncorrected by the near-infrared pyramid wavefront sensor. To address this, METIS implements dedicated focal-plane wavefront sensing strategies capable of measuring and correcting WV-driven chromatic aberrations in real time, most notably through the Asymmetric Lyot mask WaveFront sensor \citep[ALF,][]{OrbandeXivry2024,Absil2024}. Beyond instrumental mitigation, our results also indicate that improved scheduling enabled by a reliable real-time predictor of WV turbulence~\citep{Turchi2020} derived from observatory telemetry could further reduce its impact on METIS high-contrast performance.

\section{Summary and Conclusion}
\label{sec:conclusion}

In this study, we explored the variability of the differential WV column density ($\sigma_{\Sigma}$), which quantifies the strength of WV seeing, and studied its correlation with various atmospheric parameters, including PWV, seeing, wind speed, AH, RH and airmass. The analysis showed that $\sigma_{\Sigma}$ is strongly correlated with PWV, while seeing and airmass do not exhibit a significant correlation, supporting PWV as a relevant predictor for WV-induced fluctuations. A strong scatter around the identified linear trend is observed though, and cannot be attributed to other atmospheric parameters, even after exploring the influence of the vertical profiles of atmospheric turbulence and humidity. Feature importance analysis, using both Pearson correlation and RF permutation, indicates that the fringe tracker SNR is a strong contributor to the estimated strength of WV seeing, which could be at the origin of the strong scatter in the PWV vs WV seeing linear relationship. Our analysis of the power spectral density of WV seeing confirms previous studies suggesting that WV seeing follows the same Kolmogorov power spectrum as dry air turbulence. 

Our statistical characterization of the strength of WV seeing demonstrates that the majority ($\sim 95\%$) of the collected data falls below the $6~\mu$m rms threshold that was previously used to assess the impact of WV seeing on the METIS HCI performance \citep{Absil2022}. The median strength of WV seeing derived in the present study is about twice as low as the assumption used in this earlier work, resulting in an RMS wavefront error of about 15 nm RMS in the $L$-band and 175 nm RMS in the $N$-band relative to the $K$-band. While updating the full end-to-end performance analysis described in this previous paper is beyond the scope of the present work, we can still conclude that the effect of WV seeing will generally be less severe than assumed in \citep{Absil2022}, while still remaining a dominant source of HCI performance degradation, especially at $N$-band. In addition to real-time focal-plane wavefront sensing and control, reducing the impact of WV seeing will require dynamical scheduling to be implemented at the ELT to benefit from the best PWV conditions during METIS HCI observations.

\section*{Acknowledgements}
This study received funding from the Innovation in Science Pursuit for Inspired Research (INSPIRE) Fellowship, Department of Science and Technology, Government of India; the European Research Council (ERC) under the European Union’s Horizon 2020 research and innovation programme (grant agreement No 819155), and from the Wallonia-Brussels
Federation (grant for Concerted Research Actions).

\section*{Data Availability}
This study utilizes GRAVITY/VLTI fringe-tracker data for deriving $\sigma_{\Sigma}$. Vertical absolute humidity profiles obtained from the LHATPRO radiometer are available at the Paranal LHATPRO database. Atmospheric turbulence profiles from the Paranal site measured with the MASS instrument can be accessed via the Paranal MASS database. The wind-speed data from the Vaisala METEO station at Paranal can be accessed through the Paranal METEO Query Form. All datasets—including GRAVITY/VLTI fringe-tracker data and Paranal site data (LHATPRO, MASS and METEO)—are publicly accessible through the ESO Archive Portals.

Conflicts of Interest Statement: Authors declare no conflict of interest.

\bibliographystyle{mnras}
\bibliography{example}

\bsp	
\label{lastpage}
\end{document}